\documentclass[aps,pra,superscriptaddress,showpacs,floatfix,12pt]{revtex4}
\usepackage{graphicx}
\begin{document}

\preprint{}

\title{Entanglement, avoided crossings and quantum chaos in an Ising model with a tilted 
magnetic field}

\author{J. Karthik \footnote{Present Address: Department of Materials Science and Engineering, University of Illinois, Urbana-Champaign, IL 61801, U.S.A.. e-mail: kjambun2@uiuc.edu.}}
\author{ Auditya Sharma \footnote{Present Address: Department of Physics, University of California, Santa Cruz, CA 95064, U.S.A.. e-mail: sharma@physics.ucsc.edu}}
\author{ Arul Lakshminarayan \footnote{e-mail: arul@physics.iitm.ac.in}}
\affiliation{Department of Physics\\ Indian Institute of Technology Madras\\
Chennai, 600036, India.}


\begin{abstract}
We study a one-dimensional Ising model with a magnetic field and
show that tilting the field induces a transition to quantum chaos. We explore the
stationary states of this Hamiltonian to show the intimate connection between entanglement
and avoided crossings. In general entanglement gets exchanged between the states undergoing 
an avoided crossing with an overall enhancement of multipartite entanglement at the closest
 point of approach, simultaneously accompanied by diminishing two-body entanglement as measured by concurrence.
 We find that both for stationary as well as nonstationary states, nonintegrability leads to a destruction of two-body correlations and distributes entanglement more globally. 
\end{abstract}


\maketitle
\newcommand{\newc}{\newcommand}
\newc{\beq}{\begin{equation}}
\newc{\eeq}{\end{equation}}
\newc{\kt}{\rangle}
\newc{\br}{\langle}
\newc{\beqa}{\begin{eqnarray}}
\newc{\eeqa}{\end{eqnarray}}
\newc{\pr}{\prime}
\newc{\longra}{\longrightarrow}
\newc{\ot}{\otimes}
\newc{\rarrow}{\rightarrow}
\newc{\h}{\hat}
\newc{\bom}{\boldmath}
\newc{\btd}{\bigtriangledown}
\newc{\al}{\alpha}
\newc{\be}{\beta}
\newc{\ld}{\lambda}
\newc{\sg}{\sigma}
\newc{\p}{\psi}
\newc{\eps}{\epsilon}
\newc{\om}{\omega}
\newc{\mb}{\mbox}
\newc{\tm}{\times}
\newc{\hu}{\hat{u}}
\newc{\hv}{\hat{v}}
\newc{\hk}{\hat{K}}
\newc{\ra}{\rightarrow}
\newc{\non}{\nonumber}
\newc{\ul}{\underline}
\newc{\hs}{\hspace}
\newc{\longla}{\longleftarrow}
\newc{\ts}{\textstyle}
\newc{\f}{\frac}
\newc{\df}{\dfrac}
\newc{\ovl}{\overline}
\newc{\bc}{\begin{center}}
\newc{\ec}{\end{center}}
\newc{\dg}{\dagger}

\section{Introduction}

Entanglement content in the ground, thermal and time evolving states of various
spin models have been an active subject of recent research \cite{Woot,Woot02,Sougato,Nielsen,Nature,vidkit,buzek04,latvid,buzek05,laksub,casati,Ibose,bosemont}. While many of the 
studies have been on integrable spin-chains such as the transverse Ising and
the Heisenberg models, few have explored implications of nonintegrability and
quantum chaos, for e.g. \cite{laksub,casati}. Entanglement itself in these many-body systems is not characterized
by a single number, but rather is revealed in various measures of multipartite
correlations. Motivated by the rapid developments in quantum information theory \cite{NCPP}
these studies have revealed a rich phenomenology including entanglement scaling 
at zero temperature second order phase transitions and logarithmic (in number of
spins) divergence of subsystem entropy at quantum critical points \cite{Nature,vidkit}.

The relationship between quantum chaos and entanglement is a complex one and has also
been studied in various systems \cite{furuyasarkar,ArulJay,tanaka,prosen,lahiri,scott,jacquod,ghose,kus,Wein,ArulSub,BLi}. It appears that typically chaos can enhance entanglement especially of a multipartite kind. Results suggest that initially unentangled states are capable of developing large multipartite entanglement under quantum chaotic evolutions that are persistent in time \cite{laksub,scott}. However integrable evolutions can generate large entanglements for specific initial states at specific times \cite{laksub}. Early studies were
based on bipartite systems such as coupled tops and pendula where numerical
results \cite{furuyasarkar,ArulJay} and random matrix modelling showed that the entanglement as measured by a subsystem von Neumann entropy was enhanced in regions of quantum chaos \cite{ArulJay}. Later studies of many-body
systems including spin models and quantum maps showed a subtler relationship \cite{laksub,casati,kus,Wein}, but there
is significant evidence that multipartite entanglement is enhanced by quantum chaos \cite{laksub,scott,Wein}. 
Related systems are disordered spin chains wherein transitions in level statistics with increasing disorder have been shown to be correlated with decreasing two-body entanglements, for instance in \cite{Santos04}.

The mechanisms that underlie the correlations between entanglement and nonintegrability
are not entirely explored. It may simply be that the entanglement content of random
states are reflected in systems with quantum chaos. The critical requirement of operators for producing
large multipartite entanglement has been explored before \cite{Wein} and there are indications that the
random distribution of matrix elements, rather than other characteristics such as the nearest-neighbor
spacing distribution (NNSD), are needed for generating high entanglement. On the other hand such a distribution 
may not be easily generated. Also connections between localization and entanglement have been noticed earlier \cite{ArulSub,BLi}.
 
In this paper we point to connections between nonintegrability and entanglement via avoided crossings.
 One of the hallmarks of quantum chaos is level repulsion, the tendency of quantum energy levels to avoid each other \cite{Haake}. This is most clearly seen when a parameter of the system is changed. The resultant energy level dynamics has a typical behavior for quantum chaotic systems once the levels are restricted to 
the same symmetry class. There are no exact or accidental crossings, levels come close
to each other and get "scattered". The effect of this is reflected in the most 
popular diagnostic of quantum chaos, namely the nearest-neighbor spacing statistic.
Due to level repulsion this deviates from the Poisson statistics obtained for
integrable systems to the Wigner distribution typical of quantum chaos \cite{Haake}.

When two levels approach each other due to the variation of one parameter,
from a theorem of von Neumann and Wigner we know that generically they will not
become degenerate \cite{VonWig}. The avoided crossing can be very sharp or broad and may also
involve other levels. The nature of the eigenstates ``morph'' into one another
at an avoided crossing \cite{Reichl}. As avoided crossings are generically found in 
nonintegrable systems, it seems natural to look at the behavior of entanglement
at these points. This has been exploited to some extent in earlier works that seek to create
entangled states such as the $W$ or GHZ state by using the superpositions that develop at 
avoided crossings \cite{GHZW1,GHZW2,GHZW3}, we study this more systematically in a nonintegrable system.
For this same system we show that there is a close correlation of the NNSD to the extent of entanglement;
 the Poisson statistic favoring nearest neighbor
correlations and low multipartite ones, while the Wigner distribution favors large multipartite 
entanglements with low two-body correlations. We show that for time evolving states this is reflected in the
way an initially maximally entangled pair of spins evolve and share this entanglement along with that
created by the dynamics. Once again the case when there is chaos leads to a destruction of two-body
correlations alongwith enhanced multipartite entanglement.

\section{The model}

The Hamiltonian we will use in this study is
\begin{equation}
H(J,B,\theta) = J \sum_{n=1}^{L-1} \sg^z_n \sg^z_{n+1} + B \sum_{n=1}^{L}(\sin(\theta) \sg^x_n + \cos(\theta) \sg^z_n)
\end{equation}
For $\theta=0$ the magnetic field is longitudinal, the model is almost trivially 
integrable and the spectrum is highly degenerate. When $\theta=90 ^{\circ}$ the 
field is transverse and the model is still integrable thanks to the 
Jordan-Wigner transform \cite{JW,TIM} that maps the model to one of noninteracting fermions. 
This has been extensively studied both from the original statistical physics
motivation and also recently from the viewpoints of quantum information theory.
When $0<\theta <90^{\circ}$, the model is not integrable, when converted using 
the Jordan-Wigner transform the resulting fermions are interacting.  When
the magnetic field is pulsed or kicked, this has been studied as the ``kicked
Ising model'' and has also been used to study entanglement \cite{laksub}. The kicked 
Ising model is believed to be quantum chaotic for intermediate tilt angles \cite{prosen2}. However
the time dependence is not essential in introducing nonintegrability and we will
study the autonomous Hamiltonian above. It is evidently a very natural generalization of the well studied case of the transverse Ising model \cite{TIM}. We have since doing this work noticed that it has appeared in two other
recent complimentary studies \cite{casati,proszni}.

The antiferromagnetic and ferromagnetic chain spectrums are related to each other as 
\beq
\left( \prod_{i=1}^L \otimes \,\sg^y_i \right) \, H(J,B,\theta) \, \left(\prod_{i=1}^L \otimes \,\sg^y_i \right)\, =\, -H(-J,B,\theta).
\eeq
The ground state of the antiferromagnetic chain is the spin-flipped version of the most excited state
of the corresponding ferromagnetic chain. We will consider $J>0$, as we will mostly deal with the 
entire spectrum of states.
A discrete symmetry present for all angles is that of interchanging the spins at the sites $i$ and $L-i+1$ for all $i=1,\ldots L$, a ``bit-reversal'' symmetry as the field and interaction do not distinguish the spins except for the open boundary condition. If $B$ represents this reversal
\beq
B|s_1 s_2 \ldots s_N \kt = |s_N \ldots s_2 s_1 \kt, \;\;\; \left [ H, \, B \right]=0
\eeq
where $|s_i \kt$ is any single particle basis state, such as the standard ($S_z$) 
basis states we will use. Since $B^2=1$, the eigenstates can be classified as odd
or even on bit-reversal. The dimensionality of the odd subspace is half the number
of non-palindromic binary words of length $L$, while the even subspace is larger 
by the number of palindromic words of the same length.
The chain with periodic boundary conditions naturally
has a much larger symmetry group corresponding to a shift operation. For $\theta=0$
the spectrum is highly degenerate with eigenstates that can be chosen to
be product states. For the transverse Ising model, $\theta=\pi/2$, the spectrum
is much less degenerate, but the Jordan-Wigner transform converts the problem
into one of non-interacting fermions. For intermediate angles this transform
leads to a model of interacting fermions which is not solvable. 

\begin{figure}
\includegraphics[height=4in,width=3in,angle=270]{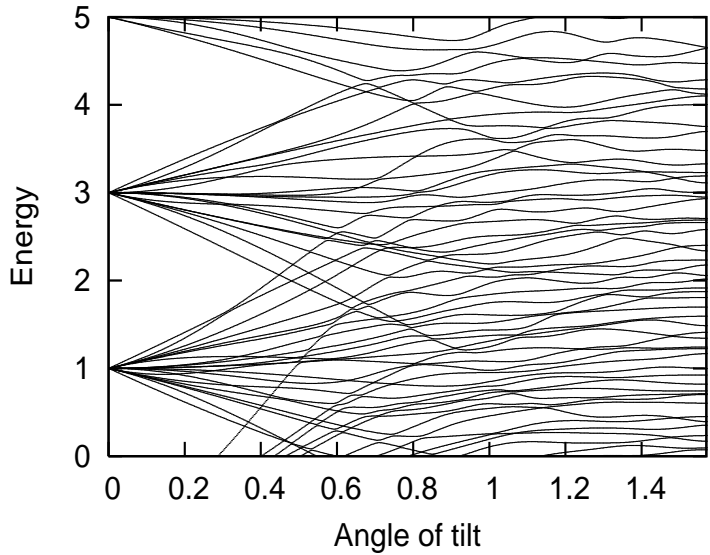}
\caption{A part of the even subspace of the energy spectrum for the case of $L=8$ spins. $J=B=1$.}
\label{spectrum}
\end{figure}

In Fig.~(\ref{spectrum}) we show a part of the even subspace of the 
energy spectrum for $L=8$ spins. 
The fanning out of the energy eigenvalues from degenerate ones is typical of
systems that lose symmetry and becomes nonintegrable. The apparent crossings
of energy levels are in fact very close avoided crossings. As the tilt angle
increases further the avoided crossings become more apparent and dense. 

\begin{figure}
\includegraphics[height=4in,width=3in,angle=270]{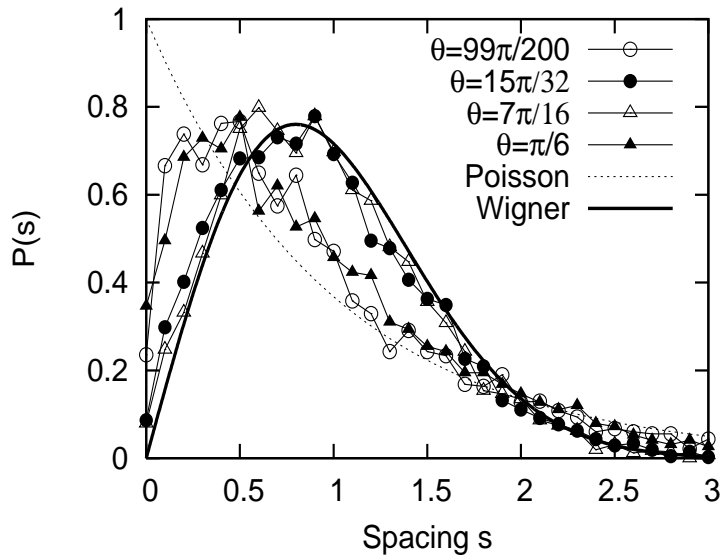}
\caption{The nearest neighbor spacing distribution of the even states of a chain with $L=13$ spins at various angles $\theta$ of tilt of the magnetic field. $J=B=1$.}
\label{nns}
\end{figure}

A standard unfolding of the spectrum in a given symmetry class is done by numerically 
fitting the staircase
function (cumulative density) to high-order polynomials and using this to 
map the energies to unfolded ones such that the mean energy spacing is unity. The NNSD is
a commonly used indicator of quantum chaos\cite{Haake,Mehta}. In Fig.~(\ref{nns}) we show the NNSD for four angles, three of them 
fairly close to a purely transversal field ($\theta/\pi=99/200,15/32,7/16$). It is 
to be noted that even for small longitudinal fields the NNSD is close to the 
Wigner distribution, corresponding to the Gaussian Orthogonal Ensemble (GOE) of Random Matrix Theory (RMT) \cite{Haake,Mehta}, given by 
\[ P_W(s)=\frac{\pi}{2} \, s\, \exp(-\pi s^2/4) \]

 The solvable pure transverse field case is not apparently completely
desymmetrized by the ``bit-reversal'' operator, degeneracies persist and therefore
we do not plot the NNSD in this case.As the tilt angle is increased further we are tending towards the solvable purely longitudinal case, with very large degeneracies. A reflection of this is seen in the NNSD deviating significantly from the GOE distribution even at $\theta=\pi/3$.
The appearance of the GOE distribution indicates a certain measure
of quantum chaos and applicability of random matrix ensembles. However it must be 
noted that the Hamiltonian we have considered is ``simple'' in that there is no 
disorder and the interaction is only nearest neighbor. We will presently quantify the 
distance of the NNSD from the Poisson one ($P_P=\exp{(-s)}$) and that of the GOE ($P_W(s)$).

\section{Entanglement in the stationary states}
We will use three measures to quantify entanglement.

\begin{enumerate}
\item Entanglement within pure states of a bipartite system can be measured
by the von Neumann entropy of the reduced density matrices. However there
are many bipartite splits, given $L$ spins. Generalized entanglement 
measures have been constructed and studied based on the entanglement 
in each of these possible splits. We will however consider only the entanglement
of the first $L/2$ spins with the remaining $L/2$ ones for $L$ even, denoted as $S_{L/2}$.
Thus if $|\psi \kt$ is an $L-$ spin pure state, 
\beq
S_{L/2}= -\mbox{tr}_{1,\ldots,L/2}\left(\rho_{1,\ldots,L/2} \, \log(\rho_{1,\ldots,L/2})\right)
\eeq
where
\beq
\rho_{1,\ldots,L/2}=\mbox{tr}_{L/2+1,\ldots,L} \left(|\psi \kt \br \psi | \right)
\eeq

\item The state of any two spins, such as nearest neighbors, is in general a
mixed state. For such a state while the entanglement can be measured as the average
entanglement of its pure-state decompositions, the existence of an
infinite number of such decompositions makes their minimization over
this set a nontrivial task. Wootters and Hill \cite{WootPRL9899} carried out
such a procedure for the case of two spin one-half (qubit) systems and
showed that a new quantity they called concurrence was a measure of
entanglement. This facilitated the study of entanglement sharing among
many qubits. If $\rho_{ij}$ is the reduced density matrix obtained by 
tracing out all spins except those at sites $i$ and $j$, and 
defining a spin-flip operator, which takes $\rho_{ij}$ to
\beq
\tilde{\rho_{ij}}=(\sigma_y\otimes\sigma_y)\rho^{*}_{ij} (\sigma_y\otimes\sigma_y),
\eeq
the concurrence of $\rho_{ij}$ is defined to be:
\beq
C(\rho_{ij})=\mbox{max}\;\;\{\sqrt{\lambda_1}-\sqrt{\lambda_2}-\sqrt{\lambda_3}-\sqrt{\lambda_4},\,0\}
\eeq
where $\lambda_i$ are the eigenvalues of the non-Hermitian matrix
$\rho_{ij} \tilde{\rho_{ij}}$. Wootters \cite{WootPRL9899} showed that the
entanglement of formation of $\rho_{ij}$ is a monotonic function of
its concurrence and that as the concurrence varies over its possible
range $[0,1]$, the entanglement of formation also varies from 0 to 1,
We will use the square of the concurrence, called tangle, summed over all possible pairs of spins
\beq
\br \tau \kt\,=\,\sum_{i<j}\, C^2(\rho_{ij})
\eeq
 as a measure of two-body correlations in the $L-$ spin state.

\item The Meyer and Wallach $Q$ measure is also being widely used as a measure
of multipartite entanglement. The geometric multipartite entanglement measure $Q$ \cite{MeyWall}, has been shown to be simply related to one-qubit purities \cite{Brennen}, which makes their calculation and interpretation straightforward. If $\rho_k$ is the
reduced densty matrix of the spin $k$ obtained by tracing out the rest of the spins then
\beq
Q(\psi)=2\left(1-\f{1}{L}\sum_{k=1}^{L}\mbox{Tr}(\rho_k^2)\right).
\eeq
\end{enumerate}

\begin{figure}
\includegraphics[height=6in,width=4in,angle=270]{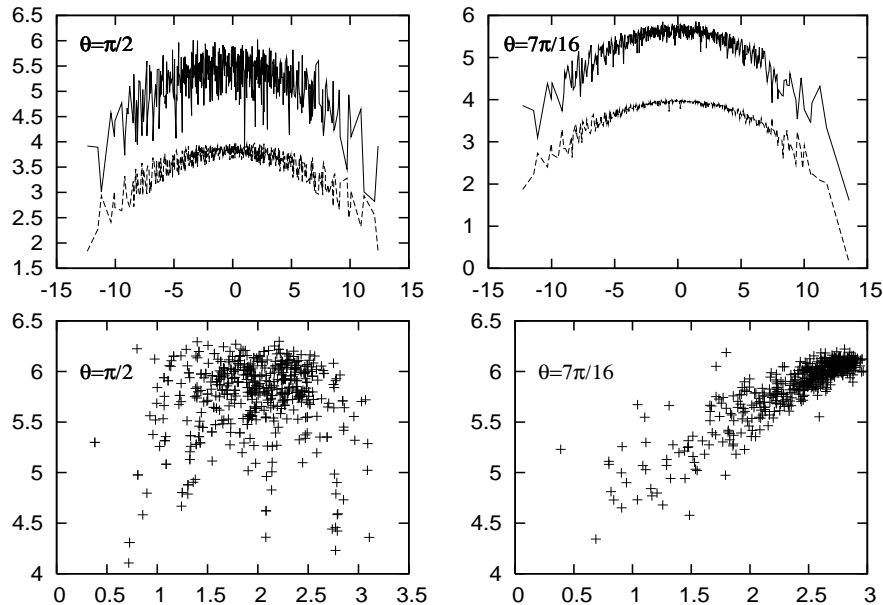}
\caption{The log of the participation ratio (solid line, top) and $4\,Q$ (dashed line, bottom) versus energy
 are shown for all eigenstates of an integrable ($\theta=\pi/2$) and a nonintegrable ($\theta=7\pi/16$ case
 in the top panels. In the bottom panels are shown, for the same cases, the entanglement measure $S_{L/2}$ versus 
the Shannon entropy $S_{sh}$ of the same states. Notice the apparent lack of correlation for the integrable case
as compared with the nonintegrable one. In all cases $L=10$ and $J=B=1$.}
\label{enrgent}
\end{figure}

In Fig.~(\ref{enrgent}) we show the entanglement content in eigenstates across a particular spectrum. We see
that there is a secular dependence of the multipartite entanglement on the energy. We connect this dependence with
other more grosser features such as localization of the eigenstates in the computational basis. Thus if 
$\psi_k=\br k |\psi \kt$ is the $k$-th eigenstate component, $k=0, \ldots,2^L-1$, we measure its spread in the
computational basis as $\log({\mbox PR})=-\log(\sum_{k}|\psi_k|^4)$ or as $S_{sh}=-\sum_k |\psi_k|^2\, \log(|\psi_k|^2)$,
i.e., the log of the participation ratio and the Shannon entropy of the states. Both these measures are
obviously single-particle basis dependent while measures of entanglement are immune to these changes.
However there is a strong correlation between measures of entanglement such as $Q$ and these grosser measures
of localization, especially in the nonintegrable regimes. It is also clear from these graphs that average
entanglement over the spectrum will be dominated by states from the central regions, away from the ground states
of the antiferromagnetic chain under consideration, as well as away from the groundstate of the ferromagnetic
chain to which the highest excited states correspond.

\begin{figure}
\includegraphics[height=6in,width=4in,angle=270]{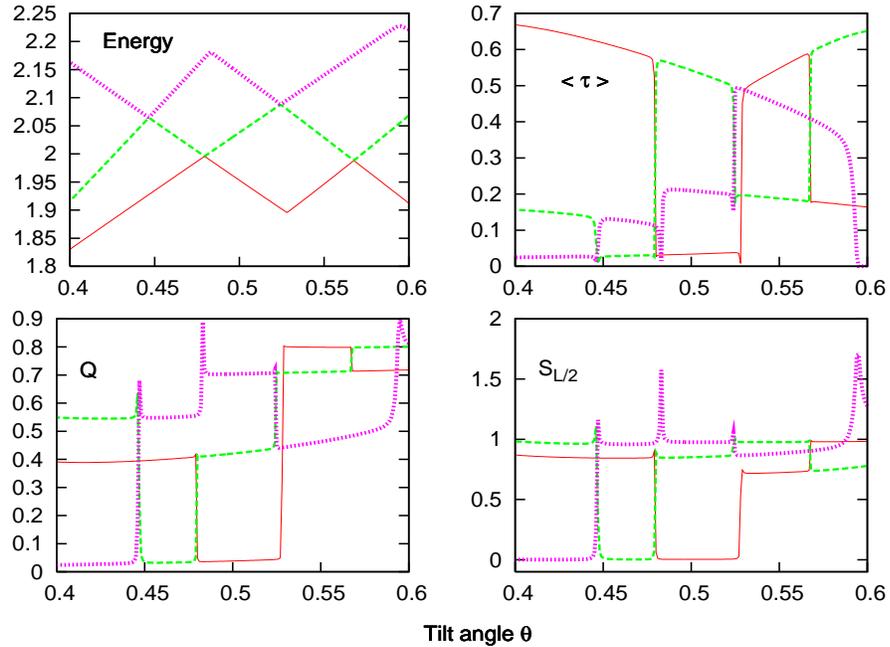}
\caption{clockwise from top-left: (1) details of three energy levels from Fig.~(\ref{spectrum}) ($J=B=1, L=8$) undergoing
avoided crossings, (2) The sum of the 2-tangles of all spin pairs, (3) The global entanglement measure $Q$ for the three states and (4) the entanglement of one-half of the chain with the other ($S_{L/2}$).}
\label{avdcross}
\end{figure}

\begin{figure}
\includegraphics[height=6in,width=4in,angle=270]{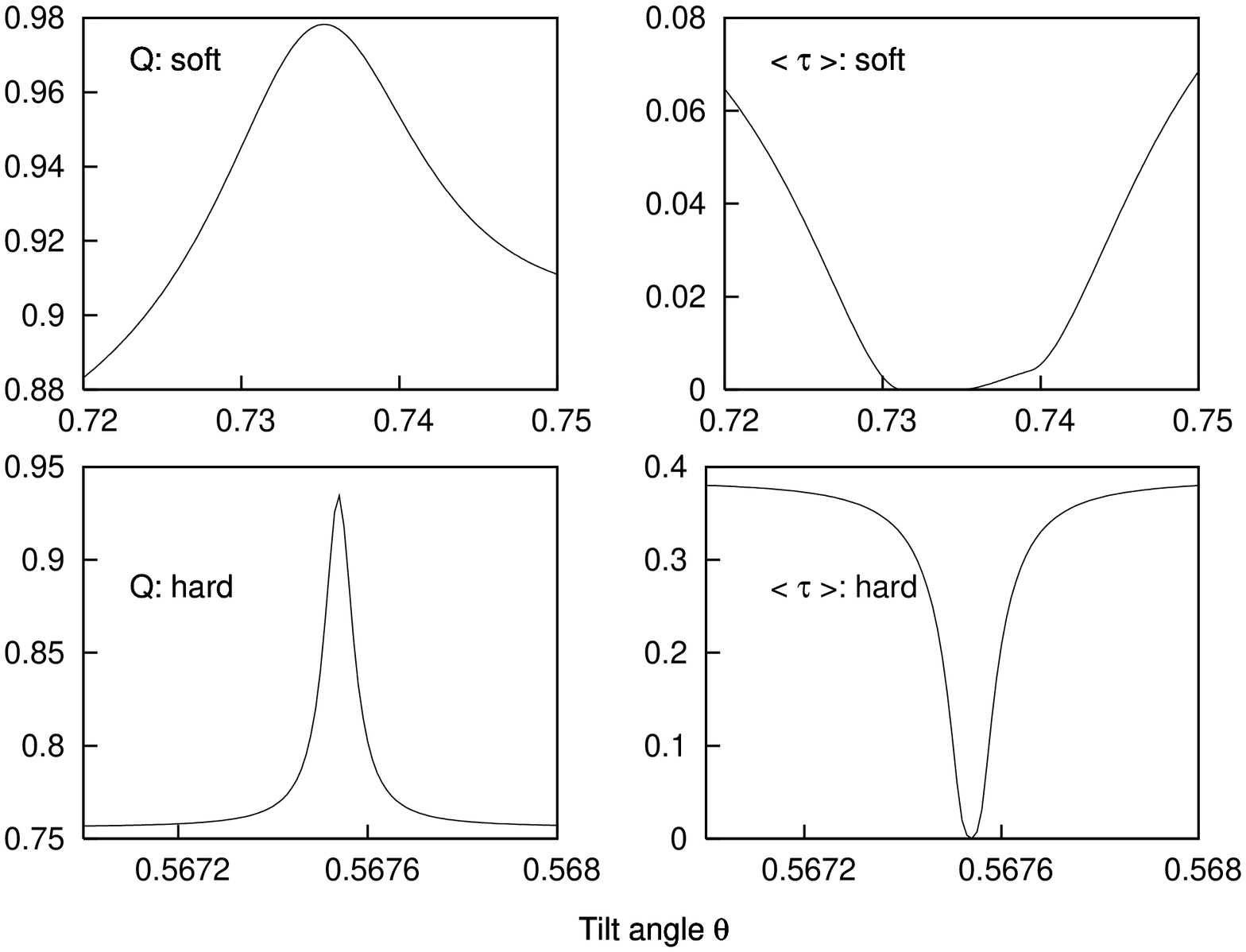}
\caption{ The average $Q$ and average total 2-tangle across two avoided crossings shown in Fig.~(\ref{avdcross}) between the lower two levels, one soft and one hard.}
\label{aventavdcrs}
\end{figure}

We now turn to how individual states' entanglement content change as the parameter $\theta$ is varied. Parametric motion of energy levels has been the subject of several studies in the past and their connection to quantum chaos is well known. 
To be specific we look at three energy levels undergoing multiple avoided crossings in a range of the parameter, some
of them sharp and some soft. Shown in Fig. (\ref{avdcross}) are the levels alongwith three measures of their entanglement,
a sum of all two-tangles, the $Q$ measure and the entropy $S_{L/2}$. The central state undergoes collisions with
its neighbors, while the other two also undergo collisions with states not shown in the figure. It is clear from all the 
measures of entanglement that at avoided crossings there is an exchange of entanglement. In particular it is interesting that the highest energy state shown has a vanishing entropy $S_{L/2}$ till it undergoes an AC with the middle level which subsequently has zero entanglement, which further gets transferred to the state with the lowest energy. Thus along with avoided crossings at which states undergo structural changes, it is reasonable that entanglement properties get exchanged as well. In the vicinity of an AC the average entanglement of the two states involved increases sharply
in ways that seem to depend on how sharp the crossing is. However at an AC while the multipartite entanglement increases, the two-body correlations decrease. This is illustrated in Fig.~(\ref{aventavdcrs}), where the total 2-tangle is seen to dip to almost zero at the point of closest approach of the AC. We note that a fine sweeping of the parameter in the vicinity of the AC is necessary to pick up this trend which is consequently not obvious in a coarser one such as in Fig.~(\ref{avdcross})

 The fact that at an AC entanglement is enhanced, and that at the point 
of closest approach the entanglement is maximized has already been used before in the context of few spins and
the GHZ state. However we see it here in the context of level dynamics of a nonintegrable system how generic it is, and that there are tradeoffs between multipartite and two-body entanglements. It maybe one mechanism by which nonintegrability
enhances multipartite entanglement. A more detailed study of entanglement at ACs is called for, especially considering that what happens at such points is a coherent superposition of orthogonal states. That superpositions will always
cause an increase of entanglement is of course untrue, however it could be true for some large class or measure 
of states that are relevant to ACs. A very recent study of entanglement due to superpositions has appeared \cite{PoP},
and maybe relevant to what happens in the vicinity of an AC.


\begin{figure}
\includegraphics[height=4in,width=3in,angle=270]{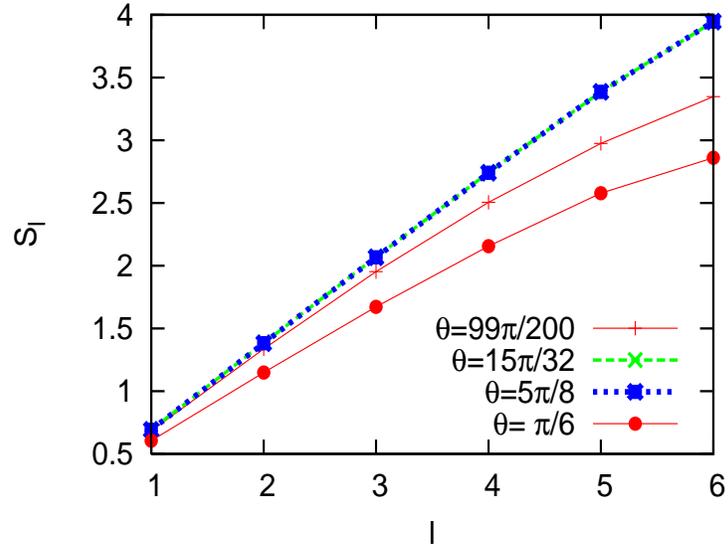}
\caption{ The subsystem entropy of $S_l$ of the first $l$ spins in a chain with $L=14$ spins ($J=B=1$) for various
tilt angles.}
\label{subsysent}
\end{figure}

We add two other numerical results that show the effect of nonintegrability on entanglement. We calculate 
the entropy $S_{l}$ of the first $l$ spins of the chain. This is known to scale as $\log(l)$ for large $l$ at quantum
critical points and tends to a constant otherwise. However it is also known that random states of relevance to 
quantum chaotic systems have subsystem entropies that scale as $\log(D)$ where $D$ is the dimensionality of the
subspace. If we are dealing with collections of qubits then $D=2^l$ and such subsystem entropies should scale
linearly with $l$, Indeed in the Ising model under consideration along with a transition to classical chaos
we observe a transition to a linear behavior. We show this in Fig.~(\ref{subsysent}) where we take data from
one hundred states at the center of the spectrum to smooth out state-to-state fluctuations. Since we cannot numerically
access very large spin chains we show calculations based on a chain of 14 spins. The entropy $S_{L/2}$ shown 
previously is a special case of $S_{l}$, but also appears to deviate from the straight lines shown in these graphs,
deviations most likely originating in the symmetry of the spin chain.
As a more direct connection to quantum chaos we quantify the distance of the spectral NNS distributions from the Poisson and the GOE distribution of RMT using the Kolmogorov-Smirnov (KS) goodness-of-fit statistic and correlate this to entanglement. The KS statistic is defined as the maximum difference between a cumulative distribution of the
data and that of a hypothesis. 
\begin{figure}
\includegraphics[height=4in,width=3in,angle=270]{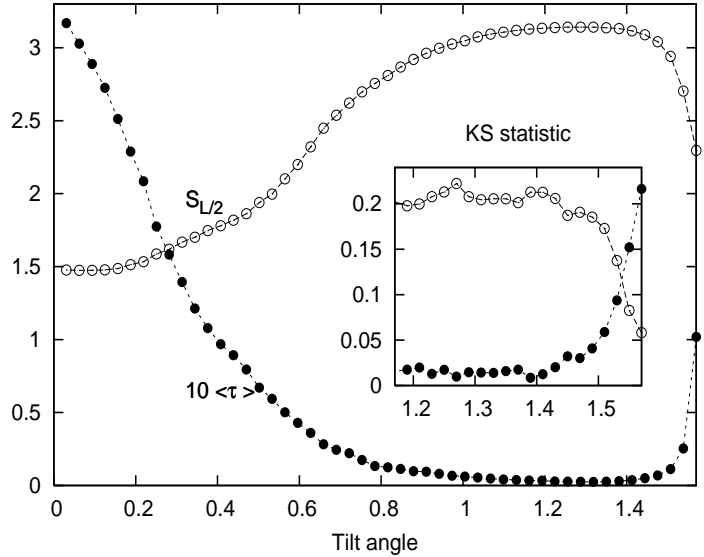}
\caption{ The subsystem entropy of $S_{L/2}$ calculated as an average over all the eigenstates, in a chain with $L=12$ spins ($J=B=1$) for a range of tilt angles of the field, the maximum corresponding to the transverse case. The inset shows the Kolmogorov-Smirnov goodness-of-fit statistic for the corresponding spectral NNS distributions, the top line corresponding to the test hypothesis being the Poisson distribution, while the lower one corresponds to the GOE distribution.}
\label{KS}
\end{figure}
As Fig.~(\ref{KS}) shows as the field gets tilted away from the transverse case, there is an increase in the 
entanglement as measured by $S_{L/2}$ while simultaneously the NNSD makes a transition to the GOE. This is seen 
for other global measures of entanglement such as the Q measure as well. This quantifies the transition from 
Poisson to GOE distributions of the NNS as well as correlates with chain entanglement properties. A similar
correlation is seen when the angle approaches zero and the there is once more a deviation from the GOE.

\section{Time Evolving States}

In this section we investigate the dynamics of entanglement for the above model. Our strategy is to start with
an initial state with exactly one maximally entangled pair and the others in an
unentangled state and study its entanglement properties under time evolution. The initial state
$|\psi\left(0\right)\rangle$ with given entanglement properties evolves to $|\psi\left(t\right)\rangle$
with a complicated distribution of entanglement. We investigate how the entanglement distribution in the chain varies with time by studying the evolution of nearest neighbur entanglement (calculated using the concurrence measure),
and the multipartite entanglement using the $Q$ measure in a chain of 10 qubit.
We consider a chain of 10 qubits with $B=1$ and $J=1$ (units chosen so that $\hbar=1$) and with the initial state:
\begin{equation}
|\psi_{e}\left(t=0\right)\rangle = \frac{1}{\sqrt2}|\left(11+00\right)111111\rangle
\end{equation}
The above state evolves with time and the plots below depict the distribution of entanglement across the chain 
as a function of time. 

\begin{figure}
\includegraphics[width=4in, height=4in]{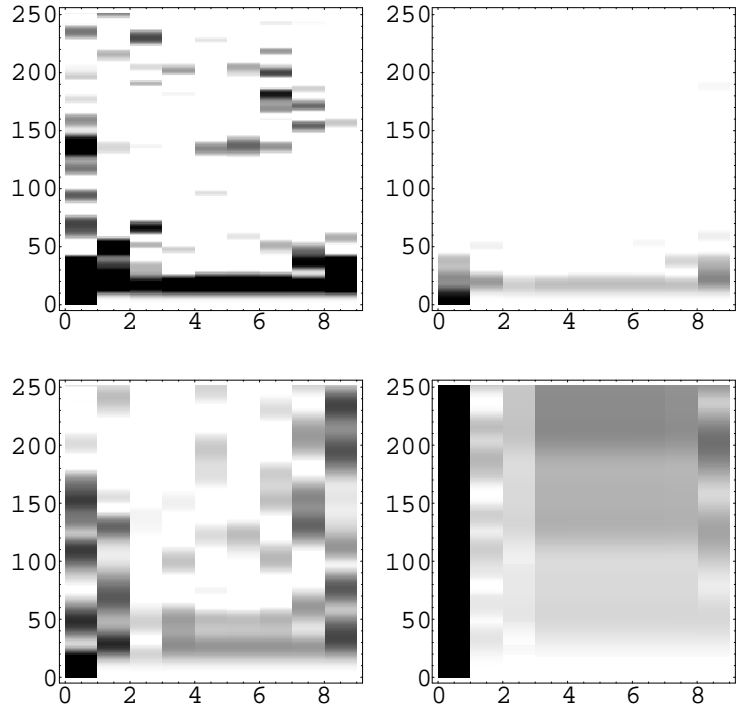}
\caption {The nearest neighbor concurrence is plotted with time for four different tilt angles ($\theta/\pi=1/2,5/12,1/3,1/6$, from left to right and top to bottom) in a chain of 10 qubits.
On the y-axis the scale is $1=1/25$ time units and $J=B=1$.}
\label{Fig: nnce}
\end{figure}

Fig.~\ref{Fig: nnce} shows the density plot of the nearest neighbor entanglement (calculated using concurrence) as a function of $l$, the location of the pair, and time, for a chain of 10 qubits with $|\psi_{e}\rangle$ as the initial state. The darkness of a region is proportional to entanglement for the given pair. We observe that entanglement is initially concentrated only between the first two qubits but as time progresses it gets distributed among other nearest neighbor pairs in the chain. We see that there is a threshold time which occurs fairly early into the evolution at which all the nearest neighbor pairs acquire entanglement. In some other studies that have investigated such transport,
such as for the Heisenberg XY spin chain, the entanglement transport has been a gradual phenomenon~\cite{Arultransport} and  tends to move as a patch across the chain with a constant velocity before being dispersed across the chain.

\begin{figure}
\includegraphics[width=4in, height=4in]{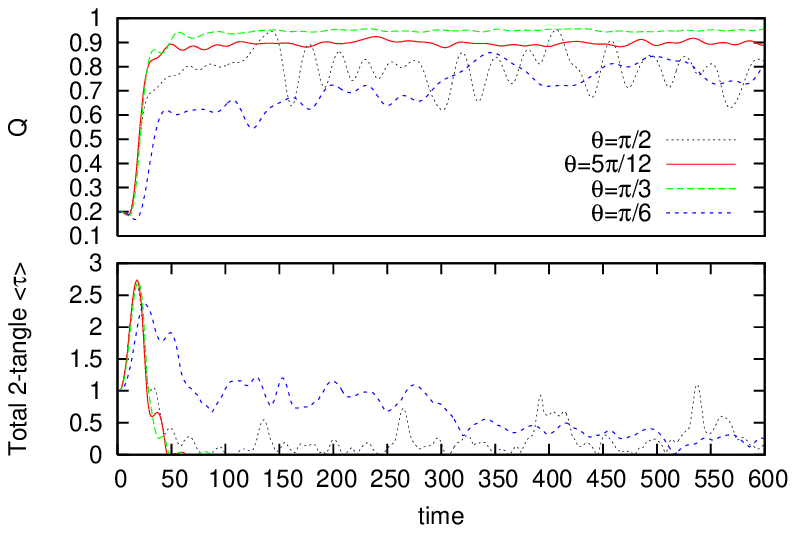}
\caption {The $Q$ Measure is plotted with time for the same four tilt angles $\theta$ as in the previous figure
for a chain of 10 qubits and $J=B=1$.}
\label{Fig:qe}
\end{figure}
However, in this case there is a sharp increase in the entanglement of the nearest neighbors and persists for a while before disappearing. The presence of this feature across the values of $\theta$ and number of spins leads us to believe that it is a robust feature  of the Hamiltonian considered. 
The dynamics of the $Q$ measure plotted in Fig. ~\ref{Fig:qe} also reveals interesting features. The sudden generation of nearest neighbor entanglement initially is accompanied by a steep rise in the $Q$ measure. Since the $Q$ measure quantifies the multipartite entanglement in the chain,  it is very likely that this steep rise is due to the nearest neighbor entanglement alone. This is followed by a drop in the nearest neighbor entanglement while the $Q$ measure remains at a large value, indicating that
as time progresses the entanglement is exclusively multipartite in nature. Further proof is obtained from the plot
of the average two-body entanglement versus time. This quantity falls with time showing that
the pairwise correlations are reduced and therefore we conclude that almost all the entanglement that survives after the initial phase is purely multipartite in nature.

We notice a close connection between the non-integrability introduced by $\theta$ and the entanglement dynamics.
Specifically, we notice that for those values of $\theta$ that result in high non-integrability for the system, the 
time taken for the quenching of two-body entanglement (concurrence) is considerably lesser. Here, the degree of non-integrability is measured by the closeness of its energy spacing distribution curve to the Wigner distribution. For $\theta=\pi/3$, the spacing distribution is very close to Wigner and we observe a quick death of two body correlations. Concurrence is seen to persist for a longer time in cases where the spacing 
distribution is significantly different from the Wigner distribution, for e.g. $\theta=\pi/2$ or $\pi/6$. 
Further, we also noticed in calculations not presented here, that average tripartite entanglement characterized by negativity also falls down faster for systems with greater non-integrability though it  persists for a somewhat longer time than the corresponding two body correlations. Therefore it seems likely that the presence of non-integrability favors the generation of higher order correlations at the expense of lower order ones.

To summarize, we have studied an Ising  spin chain with a tilted magnetic field of which the well studied one-dimensional
transverse Ising model is a special case. We have shown an intimate relationship between the spectral fluctuations of the Hamiltonian and the entanglement properties of the eigenstates. In particular we have shown that there is a larger
multipartite entanglement content when the spectral fluctuations are described by the Wigner distribution usually
found for nonintegrable systems. At the same time, such chains are low on the nearest neighbor two-body
correlations such as concurrence. We have also shown that at avoided crossings when there is generically a superposition
of ``bare" states, multipartite entanglement is enhanced while the two body concurrences are lowered. The properties
of stationary states are reflected in those of nonstationary states, and in particular we have seen that when the chain
is nonintegrable, initially present two-body entanglements are quickly quenched and are converted to multipartite entanglement while for integrable chains, significant entanglement is retained in the form of two-body concurrences.

\end{document}